\documentclass[pra,twocolumn,nofootinbib,eqsecnum,floatfix]{revtex4}
\usepackage{bm}
\usepackage{amsmath}
\input epsf
\numberwithin{equation}{section}

\def\eg{{\it e.g.~}}%

\begin{document}
\newcommand{\ltwid}{\mathrel{\raise.3ex\hbox{$<$\kern-.75em\lower1ex\hbox{$\sim$}}}}
\newcommand{\gtwid}{\mathrel{\raise.3ex\hbox{$>$\kern-.75em\lower1ex\hbox{$\sim$}}}}

\title{Excess Baggage\footnote{This article ppeared in {\sl Elementary Particles and
the Universe: Essays in honor of Murray Gell-Mann}, ed.~by J.H.~Schwarz,
Cambridge University Press, Cambridge (1991). The article has not been revised except to add an abstract, correct minor typos,  update references which were `to be published' at the time, add two footnotes concerning terminology, and restore $\hbar$ to some equations.}}

\author{James B. Hartle}
\email{hartle@physics.ucsb.edu}
\affiliation{Department of Physics,
University of California\\
Santa Barbara, CA 93106-9530 USA}

\begin{abstract}
Many advances in physics have in common that some idea which
was previously accepted as fundamental, general, and inescapable was
subsequently  seen to be consequent, special, and dispensable. The idea was not truly a general feature of the world, but only
{\it perceived}  to be general because of our special place in the universe and the limited
range of our experience. It was {\it excess baggage} which had to be jettisoned to reach a more a more general perspective. This article discusses excess baggage from the perspective of  quantum cosmology which aims at a theory of the universe's quantum initial state. We seek to answer the question `Which features  of our current theoretical framework are fundamental and which reflect our special position in the universe or its special initial condition?'  Past instances of cosmological excess baggage are reviewed such as the idea that the Earth was at the center of the universe or that the second law of thermodynamics was fundamental. Examples of excess baggage in our current understanding are the notion that measurement is central to formulating quantum mechanics, a fundamental quantum mechanical arrow of time, and the idea that a preferred time is needed to formulate quantum theory. We speculate on candidates for future excess baggage.  
\end{abstract}
\maketitle

\section{Quantum Cosmology}

It is an honor, of course, but also a pleasure for me to join in this
celebration of Murray Gell-Mann's sixtieth birthday and to address such
a distinguished audience. Murray was my teacher and more recently we
have worked together in the search for a quantum framework within which to
erect a fundamental description of the universe which would encompass
all scales -- from the microscopic scales of the elementary particle
interactions to the most distant reaches of the realm of the galaxies --
from the moment of the big bang to the most distant future that one
can contemplate.
Such a framework is needed if we accept, as we have every reason to,
that at a basic level the laws of physics are quantum mechanical.
Further, as I shall argue below, there are important features of
our observations which require such a framework for their explanation.
This application of quantum physics to the
universe as a whole has come to be called the subject of
quantum cosmology.
     
The assignment of the organizers was to speak on the topic ``Where are our
efforts leading?'' I took this as an invitation to speculate, for I think
that it is characteristic of the frontier areas of science that, while
we may know what direction we are headed, we seldom know where
we will wind up. Nevertheless, I shall not shrink from this task and
endeavor, in the brief time available, to make a few remarks on the future of
quantum cosmology.
The point of view that I shall describe owes a great deal to
my conversations with Murray.
     
One cannot contemplate the history of physics without becoming aware that
many of its intellectual advances have in common that some idea which
was previously accepted as fundamental, general, and inescapable was
subsequently  seen to
be consequent, special, and dispensable. Further, this was often for the
following reason:
The idea was not truly a general feature of the world, but only
{\it perceived}  to be
general because of our special place in the universe and the limited
range of our experience.
In fact, it arose from 
a true physical fact but one which is
a special situation in a yet more general theory.
To quote Murray himself from a talk of several years ago: ``In my field
an important new idea ..... almost always includes a negative statement,
that some previously accepted principle is unnecessary and can be
dispensed with. Some earlier correct idea .... was accompanied by
unnecessary intellectual baggage and it is now necessary to jettison
that baggage.''\cite{Gel87}
     
In cosmology it is not difficult to cite previous examples of such
transitions \cite{Hooxx}. The transition from Ptolemaic to Copernican astronomy
is certainly one. The centrality of the earth was a basic
assumption of Ptolemaic cosmology. After Copernicus, the earth
was seen not to be fundamentally central but rather one planet
among others in the solar system. The earth was in fact
distinguished, not by a law of nature, but rather by our own
position as observers of the heavens.  The idea of the central earth was
excess baggage.
     
The laws of geometry of physical space in accessible regions obey the
Euclidean laws typified by the Pythagorean theorem on right triangles.
Laws of physics prior to 1915, for example those governing the
propagation of light, incorporated Euclidean geometry as a
fundamental assumption. After Einstein's 1915 general theory of
 relativity, we see
Euclidean geometry not as fundamental, but rather as one possibility among many
others. In Einstein's theory, the geometry in the neighborhood of any body
with mass is non-Euclidean and this curvature gives a profound
geometrical explanation for the phenomena we call gravity. On the
very largest scales of cosmology, geometry is 
significantly curved and it is
the dynamics of this geometry which describes the evolution of the universe.
Euclidean geometry is the norm for
us, not because it is fundamental, but only because our observations
are mostly local  and because we happen to be living far from
objects like black holes or epochs like the big bang.
The idea of a fixed geometry was excess baggage.
     
In each of these examples there was a feature of
the current theoretical framework which was  perceived as fundamental
but which in truth was  a consequence of our particular position and
our particular time in the universe in a more general theoretical framework.
Our description was too special. There was  excess baggage
which had to be discarded to reach a more general and successful
viewpoint. Thus, in our effort to predict the future of quantum cosmology, the question naturally arises:
Which features of our {\it current} theoretical framework
reflect our special position in the universe\footnote{In a currently (2005) popular 
terminology the regularities associated with these features would be called {\it emergent}.} and which are fundamental?
Which are excess baggage?
     
We live at a special position in the universe, not so much in place, as
in time. We are late, living some ten billion years after the big
bang, a time when many interesting possibilities for physics could be
realized which are not easily accessible now.
Moreover, we live in a special universe. Ours is
a universe which is fairly smooth and simple on the largest scales,
and the evidence of the observations is that if we look earlier in time
it is smoother and simpler yet. There are simple initial conditions
which are only one very special possibility out of many we could
imagine.
     
The question I posed above can therefore be generalized: Which features of
our current theoretical framework 
are fundamental and which
reflect our special position in
the universe {\it or} our special initial conditions.\footnote{The current (2005) usage is
{\it initial condition} (singular) since there is only one wave function of the universe.}\ 
Are there natural candidates for elements of the framework which
might be generalized? Is there excess baggage? This is the question
that I would like to address today for quantum cosmology.
     
\section{Quantum Mechanics}
     
First, let us review the current quantum mechanical framework today
as it was developed in the twenties, thirties and forties, and as it
appears in most textbooks today.

It is familiar enough that there is no certainty in this world and
that therefore we must deal in probabilities. When 
probabilities are sufficiently
good we act. It wasn't certain that my plane to Los Angeles wouldn't be
blown up, but I came  because I thought that the probability was
sufficiently low.
Experimentalists can't be certain that the results of their
measurements are not in error, but they publish because they  estimate the probability
of this as low. And so on. It was the vision of classical
physics that fundamentally the world {\it was} certain and that the
use of probability was due to lack of precision. If one looked
carefully enough to start, one could with certainty predict the future.
Since the discovery of quantum mechanics some sixty years ago, we
have known that this vision is only an approximation and that
probabilities are inevitable and fundamental.  

\begin{figure}[t]
\epsfxsize=3.5in \epsfbox{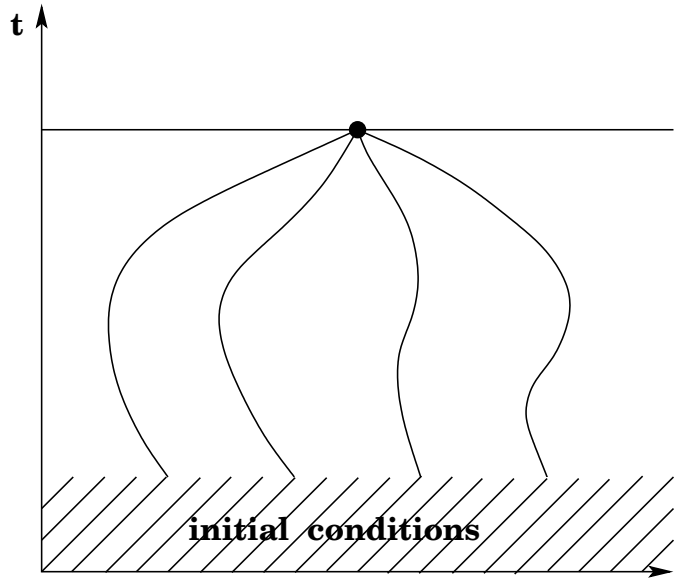}
\caption{In quantum mechanics the probability for a particle to be observed
at position $x$ at time $t$ is the absolute square of the amplitude to arrive
there.  In Feynman's sum over histories formulation, that amplitude is the
sum over all paths which are consistent with the initial conditions and
arrive at $x$ at $t$ of the amplitude for each path $\exp [iS({\rm path})/\hbar]$.
  In the Hamiltonian formulation of quantum mechanics, this sum is called
the wave function.}                                                         
\end{figure}
Indeed, the transition
from classical to quantum physics contains a good example of excess baggage.
In this case it was classical determinism.
Classical evolution was but one possibility out of many although
by far the most probable evolution in the great majority of situations
we were used to.
     
Quantum mechanics constructs probabilities according to characteristic
rules. As a simple example let us consider the the quantum mechanics of
a slowly moving particle such as an electron in a solid.
Its path in spacetime is called its history.
     
In classical mechanics the time history of the electron is a definite
path determined by the electron's initial conditions and its
equation of motion. Thus, given sufficiently precise initial conditions,
a later observation of position can yield only a single, certain,
predictable result. In quantum mechanics, given the most precise
possible initial conditions, all paths are possible and all possible
results of the observation may occur. There is only a probability
for any one of them.  The probability of an observation of position
at time $t$
yielding the value $x$, for example, is constructed as follows:
     
A complex number, called the amplitude, is assigned to each path.
This number has the form exp$[iS($path$)/\hbar]$ where the action $S$  the inertial properties of the electron together with its
interactions and $\hbar$ is Planck's constant. The amplitude to arrive at $x$ at $t$ is the sum of the
path  amplitudes over all paths which are consistent
with the initial conditions and  which end at $x$ at $t$
(Figure 1).
\begin{equation}
\left(\begin{array}{c}
\textrm{amplitude to}\\
\textrm{arrive at position}\ x\\
\textrm{at time}\ t\end{array} \right)
= \sum\limits_{\rm paths}  \exp [ i S({\rm path})/\hbar]\, .
\label{twoone}  
\end{equation}
The sum here is over all paths consistent with the initial conditions and ending
at position $x$ at time $t$.  
For example, if the position of the particle were
actually measured at a previous time but not in between, one would
sum over all paths which connect that position with $x$. 
The probability to arrive at $x$ at $t$ is the absolute square of this
amplitude.                                                        
In essence, this is Feynman's sum
over histories formulation of the rules of quantum mechanics.
     
There is another, older, way of stating these rules called the
Hamiltonian formulation of quantum mechanics. 
Here, the amplitude for observing the electron at $x$ at time $t$ is called
the wave function $\psi (x, t)$.
If, as above, we can construct the amplitude
$\psi(x,t)$ at one time,
we can also employ this construction at all other times.
The time history of the wave function gives
a kind of running summary of the probability to find the
electron at $x$.
The wave function is thus the closest analog to the classical notion
of ``state of a system at a moment of time''.  
The wave function obeys
a differential equation called the Schr\"odinger equation
\begin{equation}
i \hbar\frac{\partial \psi}{\partial t} = H \psi
\label{twotwo}
\end{equation}
where $H$ is an operator, called the Hamiltonian, whose form can
be derived from the action. It summarizes the dynamics in an
equivalent way. 
Thus, although the electron's position does not evolve according
to a deterministic rule, its wave function does.
     
Feynman's sum over histories formulation of quantum mechanics is fully
equivalent to the older Hamiltonian formulation for this case of slowly
moving particles.  One can calculate the wave function from \eqref{twoone}
 for two
nearby times and demonstrate that it satisfies the Schr\"odinger equation
\eqref{twotwo}.  One can use either formulation 
of quantum mechanics
in this and many other situations as
well.

More generally than the probability of one observation of position at one
time,
the probabilities of time {\it sequences} of observations
are of interest. Just such a sequence in needed, for example, to
check whether or not the electron is moving along a classical
path between the initial conditions and time $t$. The amplitude
for positive answers to checks of whether the electron is located
inside a sequence of position intervals is the sum of the
amplitude for a path over all paths which thread these intervals
(Figure 2).
The joint probability for the sequence of positive answers is
the square of this amplitude.

\begin{figure}[t]
\epsfxsize=3.5in \epsfbox{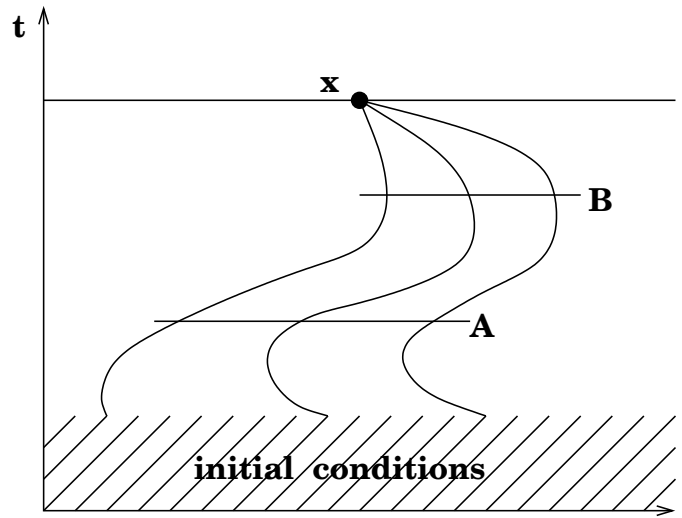}
\caption{Checks may be carried out as to whether the particle passed through
position intervals $A,\ B, \cdots$ at times in between the initial
conditions and $t$.  The amplitude to arrive at position $x$ at time $t$
with a sequence of positive answers to these checks is the sum over all
paths which meet the initial conditions and arrive at $x$ and $t$ having
threaded the intervals $A,\ B, \cdots$  The probability to arrive 
at $x$
at $t$ will depend on whether these checks are carried out {\it even
if nothing is known of the results}}.
\end{figure}

This means that the
probability of finding the particle at $x$ at $t$ depends not only
on the initial conditions, but also on what measurements were
carried out between the initial conditions and $t$.
Even if one doesn't know the results of
these measurements, the probabilities at $t$ will still depend on
whether or not they were carried out.  Thus, there are in
quantum mechanics different rules for evolution depending on
whether measurements occurred or did not. In the Hamiltonian
formulation this means that the wave function evolves by
the Schr\"odinger equation only in between measurements. At a
measurement it evolves by a different rule -- the notorious
``reduction of the wave packet''.
     
In some circumstances it makes no difference to the probabilities
whether prior measurements were carried out or not. For example,
the classical limit of quantum mechanics occurs when the initial
conditions are such that only a single path to $x$
 --- the classical one ---
contributes to the sum over histories. Then, as long as sufficiently
crude measurements are considered, it makes no difference
to the probability of $x$ at $t$ whether
they are made or not. The electron follows the classical evolution.
In such cases the classical history is said to ``decohere ''.

\section{From Bohr, To Everett, To Post-Everett}
     
The framework of quantum mechanics described in the previous section
was the starting point for the ``Copenhagen'' interpretations
of this subject.
An idea characteristic of the Copenhagen interpretations
was that there was something external to the framework
of wave function and Schr\"odinger equation which was necessary
to interpret the theory. Various expositors put this in different
ways: Bohr \cite{Boh58} spoke of alternative descriptions in terms of classical
language. Landau and Lifshitz \cite{LL58} emphasized preferred
classical observables. Heisenberg and others \cite{LB39} stressed the
importance of an external observer for whom the wave function was the
most complete summary possible of information about
an observed system.
All singled out the measurement process for a special role in the
theory.
In various ways, these authors were taking
as fundamental the manifest existence of the classical world that
we see all about us. That is, they were taking as fundamental the
existence of objects which have histories whose probabilities obey the rules of classical
probability theory and, except for the occasional intervention
of the quantum world as in a measurement, obey deterministic classical
equations of motion. This existence of a classical world, however
it was phrased, was an important part of the Copenhagen
interpretations for it was the contact with the classical world
which mandated the ``reduction of the wave packet''.

The Copenhagen pictures do not permit the application of
quantum mechanics to the universe as a whole. In the Copenhagen
interpretations the universe is always divided into two parts:
To one part quantum mechanical rules apply. To the other part
classical rules apply \cite{WPnote}.
     
It was Everett who in 1957 first suggested how to generalize the
Copenhagen framework so as to apply to
cosmology \cite{Eview}.  His idea 
(independently worked out by Murray in 1963) was to take
quantum mechanics seriously and apply it to the universe as
a whole. He started with the idea that there is one wave
function for the universe always evolving by the Schr\"odinger
equation and never reduced. Any observer would be part of the
system described by quantum mechanics, not separate from it.
Everett showed that, if the universe contains an observer,
then its activities --- measuring, recording, calculating
probabilities, etc. --- could be described in this generalized framework.
Further, he showed how the Copenhagen structure followed from
this generalization when the observer
had a memory which behaved classically and
the system under observation behaved quantum mechanically.
     
Yet, Everett's analysis was not complete. It did not explain the
manifest existence of the classical world much less the existence
of something as sophisticated as an observer with a classically
behaving memory. Classical behavior, after all, is not generally
exhibited by quantum systems. The subsequent,
post-Everett, analysis of this question
involves a synthesis of the ideas of many people. Out of many,
I might mention in particular the work of Joos and Zeh \cite{JZ85},
Zurek \cite{Zur81},
Griffiths \cite{Gri84}, and latterly Murray, Val Telegdi, and myself.
It would take us too far to attempt to review the mechanisms
by which the classical world arises but we can identify
the theoretical feature to which its origin can, for the most
part, be traced.
     
A classical world cannot be a general feature of the quantum
mechanics of the universe, for the number of states which
imply classical behavior in any sense is but a poor fraction of
the total states available to the universe. Classical behavior,
of course, can be an approximate property of a {\it particular} state as
for a particle in a wave packet whose center moves according
to the classical equations of motion. But, the universe {\it is}
in a particular state
(or a particular statistical mixture of states).  More exactly, 
particular quantum
initial conditions must be posed to make any prediction in quantum cosmology.
It is to the particular features of the initial conditions of the
universe, therefore, that we trace the origin of today's classical
world and the possibility of such information gathering and
utilizing systems as ourselves.
     
In retrospect, the Copenhagen idea that a classical world or an
observer together with
the act of measurement occupy a fundamentally
distinguished place in quantum theory can be seen to be excess
baggage. These ideas arose naturally, in part from our position
in the late universe where there {\it are}  classically behaving
objects and even observers. In part, they arose
from the necessary focus
on laboratory experiments as the most direct probe of quantum
phenomena where     there {\it is} a clear distinction between
observer and observed. However, these features of the world,
while true physical facts in these situations,
are not fundamental. Quantum mechanically, the classical world
and observers are but 
some possible systems out of many and measurements are but one possible
interaction of such systems out of many.  Both
are unlikely to exist at all in the
very early universe. They are, in a more general post-Everett
framework, {\it approximate} features
of the late universe arising from its particular quantum state.
The classical reality to which we have become so attached by
evolution is but an approximation in an entirely quantum mechanical
world made possible by specific initial conditions.
     
The originators of the Copenhagen interpretation
were correct; something beyond the wave function and
the Schr\"odinger equation {\it is} needed to interpret quantum
mechanics. But, that addition is not an external restriction of the
domain to which the theory applies. It is the initial conditions of the
universe specified within the quantum theory itself.

\section{Arrows Of Time}
     
Heat flows from hot to cold bodies, never from cold to hot. This
is the essence of the second law of thermodynamics that entropy
always increases. In the nineteenth century this law was thought
to be strict and fundamental \cite{Pai82}.  A direction of time 
was distinguished
by this law of nature. After the success of the molecular theory
of matter, with its {\it time symmetric} laws of dynamics, the increase
in entropy was seen to be {\it approximate} --- one possibility out
of many others although the overwhelmingly most probable possibility
in most situations.  A direction of time was distinguished,
not fundamentally, but rather by our position in time relative
to initial conditions of simplicity and our inability to follow
molecular motion in all accuracy. The idea of a fundamental
thermodynamic arrow of time was excess baggage.

Exchanging the thermodynamic arrow of time for simple initial
conditions might seem to be exchanging one asymmetry for another.
``Why'', one could ask, ``was the universe simple in the past and
not in the future?'' In fact, this is not a question. There is
no way of specifying that direction in time which is the past
except by calling it the direction in which the universe is
simple. Its not an arrow of time which is fundamental, but
rather the fact that the universe is simple at one end and not
at the other \cite{Haw85}.
     
The Hamiltonian formulation of quantum mechanics also possesses
a similar distinction between past and future. From the knowledge
of the state of a system at one time {\it alone}, one can predict
the future but one cannot, in general, retrodict the past \cite{aot}. 
This is an expression of causality in quantum theory. Indeed,
quantum mechanics in general prohibits the construction of history.
The two slit experiment is a famous example (Figure 3). Unless
there was a measurement, it is not just that one can't say for
certain which slit the electron went through; it is meaningless
even to assign a probability. In this asymmetry between past
and future the notion of ``state'' in quantum mechanics is
different from that of classical physics from which both future and
past can be extrapolated. In Hamiltonian quantum mechanics, the past is
the direction which is comprehensible by theory; the future
is the direction which is generally predictable.

\begin{figure}[t]
\epsfxsize=3.5in \epsfbox{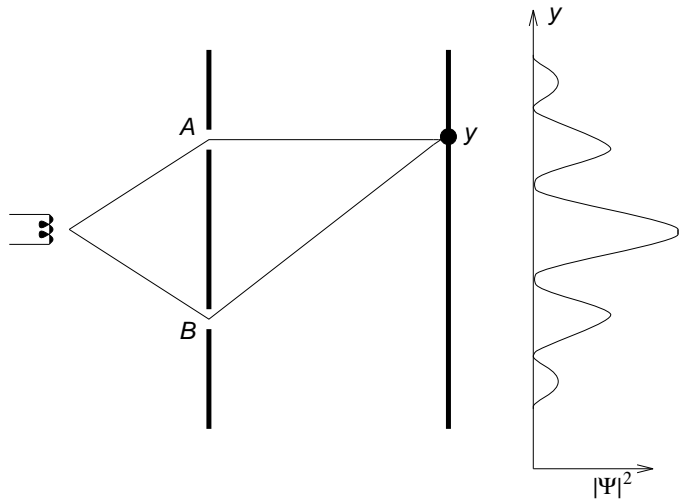}
\caption{The two slit experiment.  An electron gun at left emits an electron
traveling toward a screen with two slits,  
its progress in space recapitulating its evolution
in time.  From a knowledge of the wave function at the detecting screen
alone, it is not possible to assign a probability to whether the electron
went through the upper slit or the lower slit.}                        
\end{figure}

As long as the distinction between future and past in cosmology
is fundamental, it is perhaps reasonable to formulate a
quantum cosmology using a quantum mechanics which maintains
this same distinction as fundamental. However, it would seem more
natural if causality were an empirical
conclusion about the universe rather than a prerequisite
for formulating a theory of it.  That is, it would seem more natural
if causality were but one of many options available in
quantum mechanics, but the one appropriate for this particular
universe. For example, one might imagine a universe where both
initial {\it and} final conditions are set. The
generalization of Hamiltonian quantum mechanics necessary
to accommodate such situations is not difficult to find \cite{Harxx}. 
  In fact, it is ready to hand in the sum over histories formulation
described briefly in Section II. The rules are the same, but
both initial and final conditions are imposed on a contributing
history. In the context of such a generalization, the arrow
of time in quantum mechanics, and the associated notion of
state \cite{Unr86} would not be fundamental. Rather they would be 
features of the theory arising from the fact
that those conditions which fix our particular universe are
comprehensible at one of its ends but not at the other.

\def \G{\cal G}       

\section{Quantum Spacetime}
             
Gravity governs the structure and evolution of the universe on
the largest scales, both in space and time. To pose a quantum
theory of cosmology, therefore, we need a quantum theory of gravity.
The essence of the classical theory of gravity -- Einstein's
general relativity -- is that gravity is curved spacetime. We thus
need a quantum theory of space and time themselves in which the
geometry of spacetime will  exhibit quantum fluctuations.
     
Finding a consistent, manageable quantum theory of gravity has
been one of the goals of theoretical research for the last thirty
years. The chief problem has been seen as finding a theory which
puts out more predictions then there are input parameters.
But, in cosmology there is also a conflict with the existing
framework of quantum mechanics. These difficulties are called
``the problem of time'' \cite{pot}
     
Time plays a special and peculiar role in the familiar Hamiltonian
formulation of quantum mechanics. All observations are assumed to
be unambiguously characterized by a single moment of time and
we calculate probabilities for ``observations at one moment of
time''. Time is the only observable for which there are no
interfering alternatives (as a measurement of momentum is an
interfering alternative for a measurement of position). Time is
the sole observable not represented in the theory as an operator
but rather enters the Schr\"odinger equation as a parameter
describing evolution.

If the geometry of spacetime is fixed, external and classical, it
provides the preferred time of quantum mechanics. (More exactly,
a fixed background spacetime provides a family of timelike directions
equivalent in their quantum mechanics because of relativistic
causality.) If the geometry of spacetime is quantum mechanical --
fluctuating and without definite value -- then it cannot supply
a notion of time for quantum mechanics. There is thus a conflict
between the familiar Hamiltonian quantum mechanics with a preferred
time and a quantum theory of spacetime. This is the ``problem of
time''.
     
The usual response to this difficulty has been to keep Hamiltonian
quantum mechanics but to give up on spacetime. There are
several candidates for a replacement theory. Perhaps space and
time are separate quantum mechanically with the beautiful
synthesis  of Minkowski and Einstein
emerging only in the classical limit \cite{KA89}.  Perhaps spacetime
needs to be augmented by other, now hidden, variables which play the
role of a preferred time in quantum mechanics \cite{UWxx}. Perhaps spacetime
is simply a totally inappropriate notion fundamentally and emerges
only in the classical limit of some yet more subtle theory.
     
Each of these ideas could be right. I would like to suggest, however,
that there is another alternative. This is that the notion of a
preferred time in quantum mechanics is another case of excess
baggage. That the Hamiltonian formulation of quantum mechanics is
simply not general enough to encompass a quantum theory of
spacetime. That the present formulation of quantum mechanics is
but an approximation to a more general framework appropriate
because of our special position in the universe.
     
It is not difficult to identify that property of the late universe
which would have led to the perception that there is a preferred
time in quantum mechanics. Here, now, on all accessible
scales, from the smallest ones of the most energetic accelerators
to the largest ones of the furthest seeing telescopes, spacetime
{\it is} classical. The preferred time of quantum mechanics
reflects this true physical fact. However, it can only be an
approximate fact if spacetime is quantum mechanical.
Indeed, like all other aspects of the classical world, it must
be a property of the particular quantum initial conditions of the
universe. The present formulation of quantum mechanics thus may be
only an approximation to a more general framework appropriate
because of particular initial conditions which mandate classical
spacetime in the late universe.
     
What is this more general quantum mechanics? Feynman's sum over
histories supplies a natural candidate \cite{Tei83,Sor89,Har87}  
I shall now sketch why.
We saw earlier that
Feynman's approach was equivalent to the Hamiltonian one for particle
quantum mechanics. It is also for field theory. A little
thought about our example shows, however, that this equivalence
arises from a special property of the histories -- that they
do not double back on themselves in time. Thus, the preferred
time enters the sum over histories formulation as a restriction on
the allowed histories.
     
In a quantum theory of spacetime the histories are four dimensional
spacetimes geometries $\G$  with matter fields $\phi(x)$
living upon them. The sums over histories defining quantum amplitudes
therefore have the form:
\begin{equation}
\sum_{\G}\sum\limits_{\phi(x)} \exp \left(i\, S[{\cal G}, \phi (x)]/ \hbar \right) 
\label{fiveone}
\end{equation}
where $S$ is the action for gravity coupled to matter. The sums are
restricted by the theory of initial conditions and by the observations
whose probabilities one aims to compute.
     
The interior sum in \eqref{fiveone}
\begin{equation}
\sum_{\phi(x)} \exp \left(i\, S[{\cal G}, \phi (x)]/ \hbar \right)
\label{fivetwo}
\end{equation}
defines a usual quantum field theory in a temporarily fixed background
spacetime $\G$. The requirement that the fields be
single valued on spacetime is the analog of the requirement for particles
that the paths do not double back in time. This quantum field theory
thus possesses an equivalent Hamiltonian quantum mechanics
with the preferred time directions being those of the background $\G$.
     
When the remaining sum over $\G$ is carried out the
equivalence with any Hamiltonian formulation disappears. There
is no longer any fixed geometry, any external time, to define
the preferred time of a Hamiltonian formulation. All that is
summed over. There would be an {\it approximate} equivalence
with a Hamiltonian formulation were the initial conditions to
imply that for large scale questions in the late universe, only
a single geometry ${\hat {\G}}$ (or an incoherent sum of
geometries) contributes to the sum over geometries. For then,
\begin{equation}
\sum_{\G} \sum_{\phi(x)} \exp \left(i\, S[{\cal G}, \phi (x)]/ \hbar\right) \approx
\sum_{\phi (x) } \exp \left(i\, S[{\hat {\cal G}}, \phi (x)]/ \hbar\right)
\label{fivethree}
\end{equation}
and the preferred time can be that of the geometry ${\hat {\G}}$.
It would be in this way that the familiar formulation of quantum
mechanics emerges as an approximation to a more general sum
over histories framework appropriate to specific initial conditions
and our special position in the universe so far from the big
bang and the centers of black holes. 
The preferred time of familiar quantum mechanics would then be another example
of excess baggage.
     
In declaring the preferred time of familiar quantum mechanics excess
baggage, one is also giving up as fundamental a number of treasured
notions with which it is closely associated. These include the notion
of causality, any notion of state at a moment of time, and any
notion of unitary evolution of such states. Each of these ideas
requires a background spacetime for its definition and in a quantum
theory {\it of} spacetime there is none to be had. Such ideas,
however, would have an approximate validity  in the late universe
on those scales where spacetime is classical. Many will regard
discarding these notions as too radical a step but I think it no
less radical than discarding spacetime -- one of the most powerful
organizing features of our experience.

\section{For The Future}
     
In this discussion of the excess baggage which has been discarded to arrive
at a quantum mechanical
framework general enough to apply to the universe as a whole,
we have progressed from historical examples, through reasonable generalizations,
to topics of current research and debate. Even in the last case, I  was
able to indicate to you candidates for the necessary generalizations.
I would now like to turn to features of the theoretical framework
where this is not the case; where there may be excess baggage but for which
I have no clear candidates for the generalizations.
     
\subsection{``Initial'' Conditions} 
     
A candidate for excess baggage is
the idea that there is anything ``initial'' about the conditions which
are needed in addition to the laws of dynamics to make predictions
in the universe. Our idea that conditions are initial comes from big
bang cosmology in which the universe has a history with a simple
beginning ten and some billion years ago. The global picture of spacetime
is that there is a more or less uniform expansion everywhere from this
moment. Perhaps the very large scale structure of spacetime is very different.
Perhaps, as A. Linde has suggested \cite{Lin86} there are many beginnings. Our
local expanding universe might be just one inflationary bubble out
of many others. Perhaps, therefore, the idea that conditions should reflect
a simple state at one end of the universe is excess baggage arising
from our confinement to one of these bubbles. Conditions in
addition to dynamics would still be needed for prediction 
but they might be of a very different character \cite{univin}.
     
\subsection{Conditions and Dynamics}                                  

In the framework we have discussed so far there are three kinds of
information necessary for prediction in quantum cosmology:
First, there are the laws of dynamics
summarized by the action.  Second, there is the specification
of initial (or other) conditions. Third, there are our specific observations.
Is it fundamental that
the process of prediction be structured in this way?
     
Our focus on laboratory science, I believe, is the origin of the idea that
there is a clean separation between dynamics and conditions.
Dynamics are governed by the laws of nature; that which we are seeking,
not that which we control. By contrast, the conditions represented
in the experimental arrangement are up to us and therefore not part of the
laws of nature. This ideal 
of control is not truly realized in
practice in the laboratory, but in cosmology it is never realized. The initial
 conditions of the universe
are most definitely not up to us but must be specified in the theoretical
framework in the same law-like way as the dynamics.
A law of initial conditions has, therefore, become one of the central
objectives of quantum cosmology \cite{Har89}.
However, is it not possible
that the distinction between conditions and dynamics
is excess baggage arising from the focus on laboratory science to which
our limited resources have restricted us? Is it not possible that there
are some more general principles in a more general framework which
determine both the conditions and dynamics? Is it not
possible, as it were, that there is a principle which fixes both the
state and the Hamiltonian?
     
Recent work by Hawking \cite{Haw85}, Coleman \cite{Col88},
Giddings and Strominger \cite{GS88} among others
gives some encouragement to this point of view. They show that in
a quantum theory of spacetime which includes wormholes --- small handles
on spacetime connecting perhaps widely separated spacetime regions --
that there is a closer connection between  initial conditions and
dynamics than had been previously thought. Specifically, the
initial conditions determine the form of the Hamiltonian that we
see at energies below those on which spacetime has quantum fluctuations.
Indeed, they may determine a range of possibilities for the Hamiltonian
only one of which is realized in our large scale universe.
Dynamics and initial conditions are thereby entwined at a fundamental level.
     
\subsection{The Laws of Physics}
     
The last candidate for excess baggage that I want to discuss is the idea
that the laws of physics, and in particular laws of initial conditions,
are unique, apart from the universe, apart from
the process of their construction, and apart from us \cite{Whe77,Dav89}
Scientists, like mathematicians, proceed as though the truths of their
subjects had an independent existence. We speak, for example, of
``discovering'' the laws of nature as though there were a single set of
rules by which the universe is run with an actuality apart from this
world they govern.
     
Most honestly, the laws of physics are properties of the collective data that
we have about the universe. In the language of complexity theory, this data
is compressible. There are computational algorithms by which the data can
be stored in a shorter message. Take, for example, an observed
history of motion of
a system of classical particles. This history could be described to a given
accuracy by giving the position and momentum of each particle at a suitably
refined series of times.
However, this message can be compressed to a statement of the positions and
momenta at {\it one} time plus the equations of motion.
Some of these initial values might be specified by observation but most,
for a large system, will be specified by theory and statistically at that.
For large systems the result
is a much shorter message. Further, we find for many different systems
that the data can be compressed to the form of initial conditions plus
the {\it same} equations of motion. It is the universal character of
this extra information beyond the initial conditions
which gives the equations of motion their law-like
character. Similarly in quantum mechanics. Thus we have:

\begin{eqnarray}
{\textrm{All}\atop \textrm{Observations}} \to
\left\{\begin{array}{l}
\textrm{Some Observations}\\
               +\\
               \textrm{Laws of Dynamics}\\
               +\\
               \textrm{Laws of Initial Conditions}
               \end{array} \right .
\label{sixone}
\end{eqnarray}
The laws of physics, therefore, do not exist
independently of the data. They are properties of our data much like
``random'' or ``computable'' is a property of a number although probably
(as we shall see below)  in a less precise sense \cite{Sor83}.
                                                           
This characterization of compression is incomplete.
Any non-physicist knows that one can't just take
a list of numbers and compress them in this way. One has to take a few
courses in physics and mathematics to know what the symbols mean and
how to compute with them. That is, to the list on the right
should be appended the algorithms for numerical computation 
for practical implementation of
the theory, that part of mathematics which needed
to interpret the results, the rules of the language, etc. etc.
Yet more properly we should write: \hfil
\begin{eqnarray}
{\textrm{All} \atop \textrm{Observations}} \to 
\left\{\begin{array}{l}
\textrm{Some Observations}\\
+\\
\textrm{Laws of Dynamics}\\ 
+\\ 
\textrm{Laws of Initial Conditions}\\
+\\
\textrm{Algorithms for Calculation}\\
+\\
\textrm{Mathematics}\\
+\\
\textrm{Language}\\
+\\
\textrm{Culture}\\
+\\
\textrm{Specific Evolutionary History}\\
+ \cdots \end{array}\right.
\label{sixtwo}
\end{eqnarray}
What confidence do we have that our data can always be compressed
in this way?
I have discussed the possibility that a
clear division between initial conditions and equations of motion may
be excess baggage. Bob Geroch and I have discussed similar questions
for the division between algorithms and the rest of physical 
theory \cite{GH86}.
     
However it is subdivided,
what guarantee do we have that the resulting theory will be unique,
independent of its process of construction, independent of the
specific data we have acquired?
Very little it seems. As we move down the list on the right we encounter
more and more items which seem particular to our specific history as
a collectivity of observers
and to our specific data.
It is an elementary observation that there are always many theories
which will fit a given
set of data just as there are many curves which will interpolate between
a finite set of points. 
Further, when the data is probabilistic as it is in quantum theory,
there is always the possibility of arriving at different theories whatever
criteria are used to distinguish them. 

Beyond this, however, what confidence do we have that different groups of
observers, with different histories, with growing but different sets of
specific data, will, in the fullness of time, arrive at the same fundamental
theory?  I do not mean to suggest that the theories might vary because they
are {\it consequences} of specific history, for that is not science.  Rather the
question is whether there observationally indistinguishable theories which are
different because of the processes of their discovery.
     
There may be specifically cosmological reasons to expect non-uniqueness
 in theories of initial conditions.
A theory of initial conditions, for example, must
be simple enough that it can be stored within the universe. If the initial
conditions amounted to some particular 
complex specification of the state
of all matter this would not be possible. The act of constructing
theories may limit our ability to find them. The gravitational effect of
moving a gram of matter on Sirius by one centimeter in an unknown direction
is enough to ruin our ability to extrapolate classically the motions of
a mole of gas particles beyond a few seconds into the past \cite{Zeh84}.  In
view of this there must be many theories of initial conditions
rendered indistinguishable simply by our act of constructing them.
     
It would be interesting, I think, to have a framework which dispensed
with the excess baggage that the laws of physics were
separate from our observations of the universe,
a framework in which the inductive process of constructing
laws about the universe was
described in it, and in which our theories were seen as but one
possibility among many.
     
\section{Conclusion}
     
The assignment of the organizers was not just to speak on the subject ``Where
are our efforts leading?''.  They also wanted to know ``In fifty or one
hundred years time how do you think today's efforts will appear?''.
I have been bold enough to try their first question, I shall not be
foolish enough to essay the second. I shall, however, offer a hope, and
that is this: That in the future this might be seen a the time when scientists
began to take seriously the idea that it was important to consider
the universe as a whole and science as a unity, the time when they
began to take seriously the search for a law of how the universe started,
began to work out its implications for science generally, and began to
discard the remainder of our excess baggage.
     
\acknowledgments

The author is grateful for the support of the John Simon Guggenheim 
Foundation and
the National Science Foundation (NSF grant PHY 85-06686) during the 
preparation of this essay.  He would also like to thank the Department
of Applied Mathematics and Theoretical Physics, Cambridge University for
hospitality while it was being written.


\begin{thebibliography}{99}
\bibitem{univin}  
Few of the proposals currently under investigation as candidates
for the conditions of the universe genuinely have the character of
``initial'' conditions in the sense of posing conditions at a definite moment 
of time.
Indeed, characteristically, as described in Section V, the notion of a
single time breaks down in the early universe.  Rather these proposals fix
something like the quantum state of the universe for all times.  But also
typically these proposals imply simplicity in the early universe 
and a global spatial structure which consists of a single inflationary
bubble. The ``no boundary" proposal \cite{Haw82} is a case in point.
The point of the remarks in section VI.A  is that there may
be other possibilities.
\end{thebibliography}
\end{document}